\documentclass[aps,prd,preprint,superscriptaddress]{revtex4-1}
\usepackage[utf8]{inputenc}
\usepackage{gensymb}
\usepackage{graphicx,color}
\usepackage[section]{placeins}
\usepackage{float}
\usepackage{hyperref}
\usepackage[T1]{fontenc}
\usepackage{footmisc}
\usepackage{csvsimple}
\usepackage{longtable}

\begin{document}

\title{Measurement of the Multi-Neutron $\bar{\nu}_{\mu}$ Charged Current Differential Cross Section at Low Available Energy on Hydrocarbon}

\newcommand{\Rutgers}{Rutgers, The State University of New Jersey, Piscataway, New Jersey 08854, USA}
\newcommand{\Hampton}{Hampton University, Dept. of Physics, Hampton, VA 23668, USA}
\newcommand{\Dortmund}{Institute of Physics, Dortmund University, 44221, Germany }
\newcommand{\Otterbein}{Department of Physics, Otterbein University, 1 South Grove Street, Westerville, OH, 43081 USA}
\newcommand{\JMU}{James Madison University, Harrisonburg, Virginia 22807, USA}
\newcommand{\Florida}{University of Florida, Department of Physics, Gainesville, FL 32611}
\newcommand{\UCIrvine}{Department of Physics and Astronomy, University of California, Irvine, Irvine, California 92697-4575, USA}
\newcommand{\CBPF}{Centro Brasileiro de Pesquisas F\'{i}sicas, Rua Dr. Xavier Sigaud 150, Urca, Rio de Janeiro, Rio de Janeiro, 22290-180, Brazil}
\newcommand{\PUCP}{Secci\'{o}n F\'{i}sica, Departamento de Ciencias, Pontificia Universidad Cat\'{o}lica del Per\'{u}, Apartado 1761, Lima, Per\'{u}}
\newcommand{\INRM}{Institute for Nuclear Research of the Russian Academy of Sciences, 117312 Moscow, Russia}
\newcommand{\Jlab}{Jefferson Lab, 12000 Jefferson Avenue, Newport News, VA 23606, USA}
\newcommand{\Pittsburgh}{Department of Physics and Astronomy, University of Pittsburgh, Pittsburgh, Pennsylvania 15260, USA}
\newcommand{\Guanajuato}{Campus Le\'{o}n y Campus Guanajuato, Universidad de Guanajuato, Lascurain de Retana No. 5, Colonia Centro, Guanajuato 36000, Guanajuato M\'{e}xico.}
\newcommand{\Athens}{Department of Physics, University of Athens, GR-15771 Athens, Greece}
\newcommand{\Tufts}{Physics Department, Tufts University, Medford, Massachusetts 02155, USA}
\newcommand{\WM}{Department of Physics, William \& Mary, Williamsburg, Virginia 23187, USA}
\newcommand{\FNAL}{Fermi National Accelerator Laboratory, Batavia, Illinois 60510, USA}
\newcommand{\Purdue}{Department of Chemistry and Physics, Purdue University Calumet, Hammond, Indiana 46323, USA}
\newcommand{\MCLA}{Massachusetts College of Liberal Arts, 375 Church Street, North Adams, MA 01247}
\newcommand{\UMD}{Department of Physics, University of Minnesota -- Duluth, Duluth, Minnesota 55812, USA}
\newcommand{\Northwestern}{Northwestern University, Evanston, Illinois 60208}
\newcommand{\UNI}{Facultad de Ciencias F\'{i}sicas, Universidad Nacional Mayor de San Marcos, CP 15081, Lima, Per\'{u}}
\newcommand{\Rochester}{Department of Physics and Astronomy, University of Rochester, Rochester, New York 14627 USA}
\newcommand{\Austin}{Department of Physics, University of Texas, 1 University Station, Austin, Texas 78712, USA}
\newcommand{\USM}{Departamento de F\'{i}sica, Universidad T\'{e}cnica Federico Santa Mar\'{i}a, Avenida Espa\~{n}a 1680 Casilla 110-V, Valpara\'{i}so, Chile}
\newcommand{\Geneva}{University of Geneva, 1211 Geneva 4, Switzerland}
\newcommand{\Chicago}{Enrico Fermi Institute, University of Chicago, Chicago, IL 60637 USA}
\newcommand{\hired}{}
\newcommand{\OregonState}{Department of Physics, Oregon State University, Corvallis, Oregon 97331, USA}
\newcommand{\oxford}{Oxford University, Department of Physics, Oxford, OX1 3PJ United Kingdom}
\newcommand{\umiss}{University of Mississippi, Oxford, Mississippi 38677, USA}
\newcommand{\upenn}{Department of Physics and Astronomy, University of Pennsylvania, Philadelphia, PA 19104}
\newcommand{\AMU}{Department of Physics, Aligarh Muslim University, Aligarh, Uttar Pradesh 202002, India}
\newcommand{\wroclaw}{University of Wroclaw, plac Uniwersytecki 1, 50-137 Wroa\l{}aw, Poland}
\newcommand{\Mohali}{Department of Physical Sciences, IISER Mohali, Knowledge City, SAS Nagar, Mohali - 140306, Punjab, India}
\newcommand{\CINVESTAV}{Departamento de Fisica Col. San Pedro Zacatenco, 07360 Mexico, DF, Av. Instituto PolitÃ©cnico Nacional, Mexico}
\newcommand{\york}{York University, Department of Physics and Astronomy, Toronto, Ontario, M3J 1P3 Canada}
\newcommand{\ND}{Department of Physics and Astronomy, University of Notre Dame, Notre Dame, Indiana 46556, USA}
\newcommand{\ICL}{The Blackett Laboratory,  Imperial College London,  London SW7 2BW, United Kingdom}
\newcommand{\warwick}{Department of Physics, University of Warwick, Coventry, CV4 7AL, UK}
\newcommand{\qmul}{G O Jones Building, Queen Mary University of London, 327 Mile End Road, London E1 4NS, UK}

\newcommand{\mascencioThanks}{Now at Iowa State University, Ames, IA 50011, USA}
\newcommand{\amitbashyalThanks}{Now at  High Energy Physics Department, Argonne National Lab, 9700 S Cass Ave, Lemont, IL 60439}
\newcommand{\ricfregianThanks}{now at Department of Physics and Astronomy, University of California at Davis, Davis, CA 95616, USA}
\newcommand{\anfilkinsThanks}{now at Syracuse University, Syracuse, NY 13244, USA}
\newcommand{\finerThanks}{Now at Los Alamos National Laboratory, Los Alamos, New Mexico 87545, USA}
\newcommand{\kleykampThanks}{now at Department of Physics and Astronomy, University of Mississippi, Oxford, MS 38677}
\newcommand{\adrianThanks}{Now at Department of Physics, Drexel University, Philadelphia, Pennsylvania 19104, USA}
\newcommand{\bamThanks}{Now at University of Minnesota, Minneapolis, Minnesota 55455, USA}
\newcommand{\byaeggyThanks}{Now at Department of Physics, University of Cincinnati,  Cincinnati, Ohio 45221, USA}
\newcommand{\lazazuetareyesThanks}{now at Syracuse University, Syracuse, NY 13244, USA}

\author{A.~Olivier}                       \affiliation{\ND}  \affiliation{\Rochester}
\author{T.~Cai}                           \affiliation{\york}  \affiliation{\Rochester}
\author{S.~Akhter}                        \affiliation{\AMU}
\author{Z.~~Ahmad~Dar}                    \affiliation{\WM}  \affiliation{\AMU}
\author{V.~Ansari}                        \affiliation{\AMU}
\author{M.~V.~Ascencio}\thanks{\mascencioThanks}  \affiliation{\PUCP}
\author{M.~Sajjad~Athar}                  \affiliation{\AMU}
\author{A.~Bashyal}\thanks{\amitbashyalThanks}  \affiliation{\OregonState}
\author{A.~Bercellie}                     \affiliation{\Rochester}
\author{M.~Betancourt}                    \affiliation{\FNAL}
\author{J.~L.~Bonilla}                    \affiliation{\Guanajuato}
\author{A.~Bravar}                        \affiliation{\Geneva}
\author{H.~Budd}                          \affiliation{\Rochester}
\author{G.~Caceres}\thanks{\ricfregianThanks}  \affiliation{\CBPF}
\author{G.A.~D\'{i}az~}                   \affiliation{\Rochester}
\author{J.~Felix}                         \affiliation{\Guanajuato}
\author{L.~Fields}                        \affiliation{\ND}
\author{A.~Filkins}\thanks{\anfilkinsThanks}  \affiliation{\WM}
\author{R.~Fine}\thanks{\finerThanks}     \affiliation{\Rochester}
\author{A.M.~Gago}                        \affiliation{\PUCP}
\author{P.K.Gaur}                         \affiliation{\AMU}
\author{S.M.~Gilligan}                    \affiliation{\OregonState}
\author{R.~Gran}                          \affiliation{\UMD}
\author{E.Granados}                       \affiliation{\Guanajuato}
\author{D.A.~Harris}                      \affiliation{\york}  \affiliation{\FNAL}
\author{A.L.~Hart}                        \affiliation{\qmul}
\author{D.~Jena}                          \affiliation{\FNAL}
\author{S.~Jena}                          \affiliation{\Mohali}
\author{J.~Kleykamp}\thanks{\kleykampThanks}  \affiliation{\Rochester}
\author{A.~Klustov\'{a}}                  \affiliation{\ICL}
\author{M.~Kordosky}                      \affiliation{\WM}
\author{D.~Last}                          \affiliation{\upenn}
\author{A.~Lozano}\thanks{\adrianThanks}  \affiliation{\CBPF}
\author{X.-G.~Lu}                         \affiliation{\warwick}  \affiliation{\oxford}
\author{S.~Manly}                         \affiliation{\Rochester}
\author{W.A.~Mann}                        \affiliation{\Tufts}
\author{C.~Mauger}                        \affiliation{\upenn}
\author{K.S.~McFarland}                   \affiliation{\Rochester}
\author{B.~Messerly}\thanks{\bamThanks}   \affiliation{\Pittsburgh}
\author{O.~Moreno}                        \affiliation{\WM}  \affiliation{\Guanajuato}
\author{J.G.~Morf\'{i}n}                  \affiliation{\FNAL}
\author{D.~Naples}                        \affiliation{\Pittsburgh}
\author{J.K.~Nelson}                      \affiliation{\WM}
\author{C.~Nguyen}                        \affiliation{\Florida}
\author{V.~Paolone}                       \affiliation{\Pittsburgh}
\author{G.N.~Perdue}                      \affiliation{\FNAL}  \affiliation{\Rochester}
\author{C.~Pernas}                        \affiliation{\WM}
\author{K.-J.~Plows}                      \affiliation{\oxford}
\author{M.A.~Ram\'{i}rez}                 \affiliation{\upenn}  \affiliation{\Guanajuato}
\author{H.~Ray}                           \affiliation{\Florida}
\author{N.~Roy}                           \affiliation{\york}
\author{D.~Ruterbories}                   \affiliation{\Rochester}
\author{H.~Schellman}                     \affiliation{\OregonState}
\author{C.~J.~Solano~Salinas}             \affiliation{\UNI}
\author{M.~Sultana}                       \affiliation{\Rochester}
\author{V.S.~Syrotenko}                   \affiliation{\Tufts}
\author{E.~Valencia}                      \affiliation{\Guanajuato}  \affiliation{\WM}
\author{N.H.~Vaughan}                     \affiliation{\OregonState}
\author{A.V.~Waldron}                     \affiliation{\qmul}  \affiliation{\ICL}
\author{B.~Yaeggy}\thanks{\byaeggyThanks}  \affiliation{\USM}
\author{L.~Zazueta}\thanks{\lazazuetareyesThanks}  \affiliation{\WM}

\collaboration{The MINER$\nu$A Collaboration}
\noaffiliation
\date{\today}

\begin{abstract}
Neutron production in antineutrino interactions can lead to bias in energy reconstruction in neutrino oscillation experiments, but these interactions have rarely been studied.  MINERvA previously studied neutron production at an average antineutrino energy of $\sim$ 3~GeV in 2016 and found deficiencies in leading models.  In this paper, the MINERvA 6~GeV average antineutrino energy data set is shown to have similar disagreements.  A measurement of the cross section for an antineutrino to produce two or more neutrons and have low visible energy is presented as an experiment-independent way to explore neutron production modeling.  This cross section disagrees with several leading models' predictions.  Neutron modeling techniques from nuclear physics are used to quantify neutron detection uncertainties on this result.
\end{abstract}
\maketitle

\section{Introduction}
Accurately modeling neutron production by antineutrinos is critical to predicting energy reconstruction bias in current and future accelerator-based neutrino oscillation experiments.  Water Cherenkov-based experiments like T2K \cite{T2K} and HyperK \cite{HyperK} rely on charged current quasielastic (CCQE) kinematics to reconstruct antineutrino energy.  However, an antineutrino interaction on a correlated nucleon pair, a process known as a two-particle two-hole (2p2h) interaction, can easily be confused with a CCQE interaction and bias antineutrino energy reconstruction \cite{2p2hEnergyReconstruction}.  Experiments that use calorimetric energy reconstruction like NOvA \cite{NOvA} and DUNE \cite{DUNETDR} must account for as much of the energy produced in antineutrino interactions as possible to minimize bias in their energy measurements.  Previous studies \cite{CamilloNeutronPaper} \cite{FriedlandNeutronPaper} \cite{SuperFGDRecoPaper} have shown that leaving neutrons out of calorimetric energy reconstruction biases measurements of key oscillation parameters.

Current accelerator-based neutrino oscillation experiments correct for these effects using (anti)neutrino interaction generators \cite{GENIE} \cite{NEUT} \cite{NuWro}.  However, the MINERvA experiment has published evidence of discrepancies in the details of neutron simulations from GENIE \cite{MirandaThesisPaper}, one of the most commonly used generators, that have yet to be explained.  The few neutron-related measurements that have been made with GeV-scale antineutrinos so far \cite{SNONeutronMultiplicity} \cite{SuperKNeutronMultiplicity} have not published cross sections, so it is not straightforward to use these results to tune neutron modeling in oscillation experiments.  The ANNIE experiment \cite{ANNIEPhaseI} is designed to measure the differential cross section as a function of neutron multiplicity and plans to make this measurement for antineutrino interactions on water.

\begin{figure}[bhtp]
        \centering
        \includegraphics[scale=0.4]{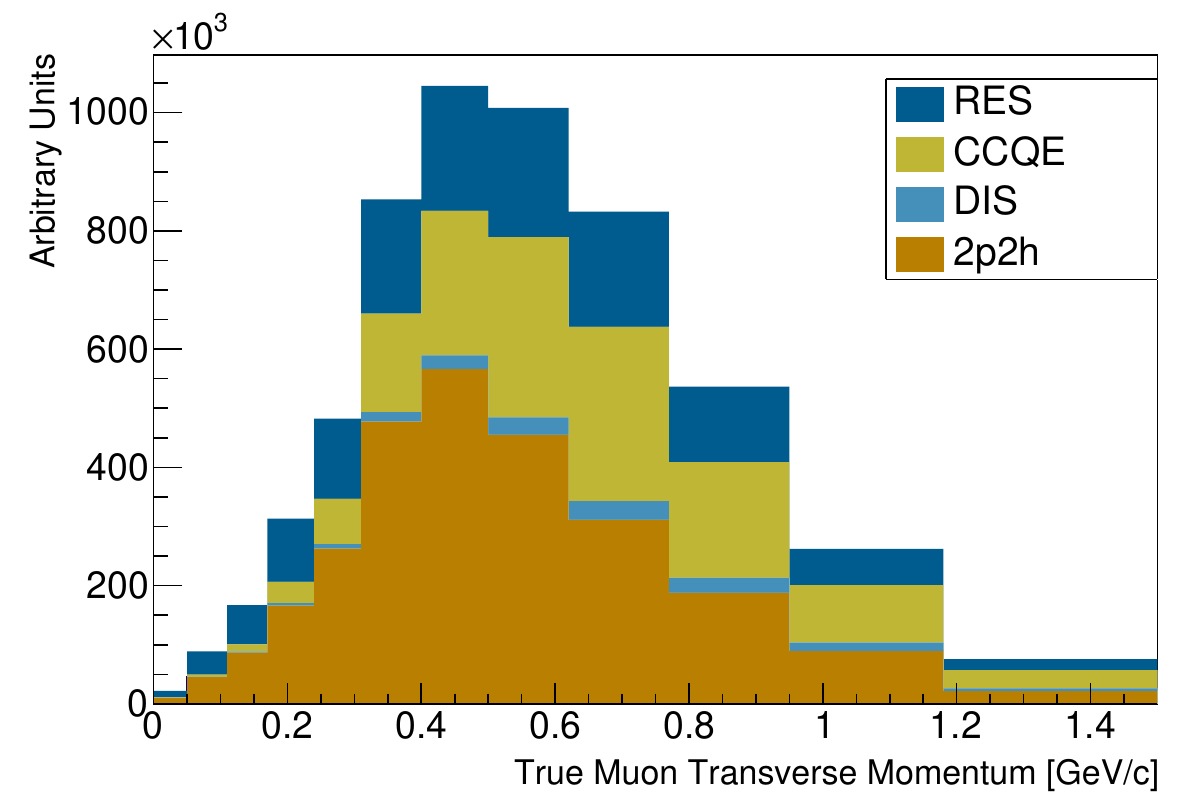}
        \caption{Multi-neutron antineutrino interactions with $E_{available}$ less than 100~MeV predicted by the main MINERvA Monte Carlo model, MnvTunev1, stacked by the interaction mode GENIE used to produce them.  2p2h and QE processes dominate across the full $p_{T \mu}$ range studied.}
        \label{fig:processBreakdown}
    \end{figure}

This paper presents the cross section for two or more neutrons to be produced in a $\bar{\nu}_{\mu}$ charged current interaction on polystyrene (CH) with visible hadronic energy (an estimator for a model quantity called $E_{available}$ defined at the end of the next section) less than 100~MeV and differential in muon transverse momentum, $p_{T \mu}$.  The cross section measured should be useful for constraining neutron production models for NOvA oscillation results and comparison to anticipated cross section measurements in T2K's new scintillator-based SuperFGD detector \cite{SuperFGDRecoPaper}.  This MINERvA sample is predicted to be particularly rich in 2p2h interactions as illustrated in Fig. \ref{fig:processBreakdown}.  It can potentially be used to constrain final state interaction (FSI) modes that turn CCQE and resonance interactions into neutron-rich final states.

\section{The MINERvA Experiment}
The MINERvA detector consists of a fine-grained hydrocarbon scintillator-based tracker deployed upstream of the MINOS near detector \cite{NIMPaper}.  It was exposed to antineutrinos from the NuMI beam \cite{NuMIPaper} at Fermilab with an average energy of 6~GeV.  The MINERvA tracker is segmented into planes of 17-mm-thick scintillator strips with three complementary orientations.  It produces detailed images of charged particles produced by (anti)neutrino interactions, and information provided by the MINOS near detector gives precise reconstruction of muon energy and muon charge sign.  Energy near the muon track, including muon-induced delta rays or bremsstrahlung photons, is also categorized as leptonic activity, but it is not used when estimating muon momentum.  All other energy is assumed to be hadronic.

\begin{figure}[bhtp]
  \centering
  \includegraphics[scale=0.4]{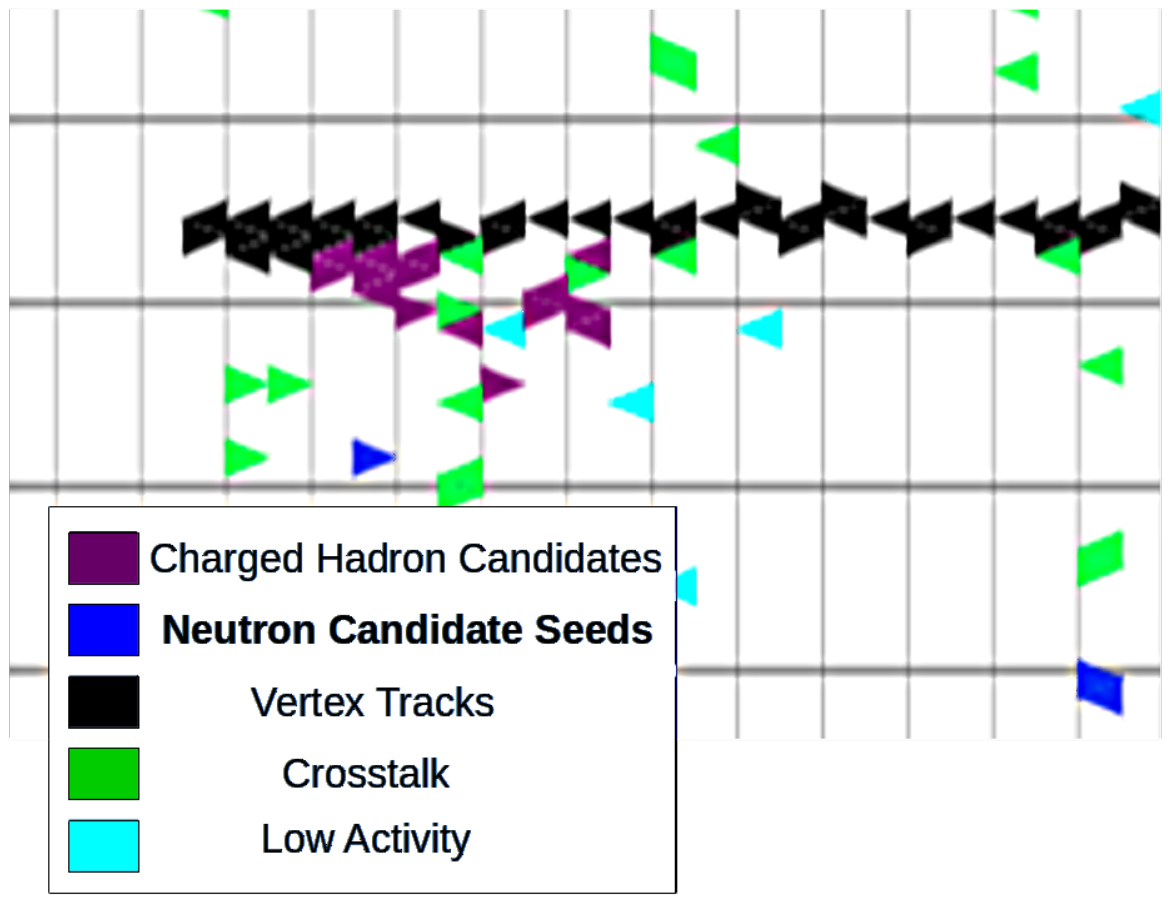}
  \caption{Illustration of energy deposits excluded from neutron reconstruction in a simulated MINERvA antineutrino interaction.  Only one of three views is shown.  Each triangle is a single scintillator strip with energy above threshold that can be thought of as a pixel in MINERvA.  The beam travels roughly along the horizontal axis.  Black strips were originally part of the vertex interaction system.  Purple strips were recursively added to the vertex region.  Dark blue strips ultimately became neutron candidates.}
  \label{fig:connectedClusterFilter}
\end{figure}

\begin{figure}[bhtp]
  \centering
  \includegraphics[scale=0.4]{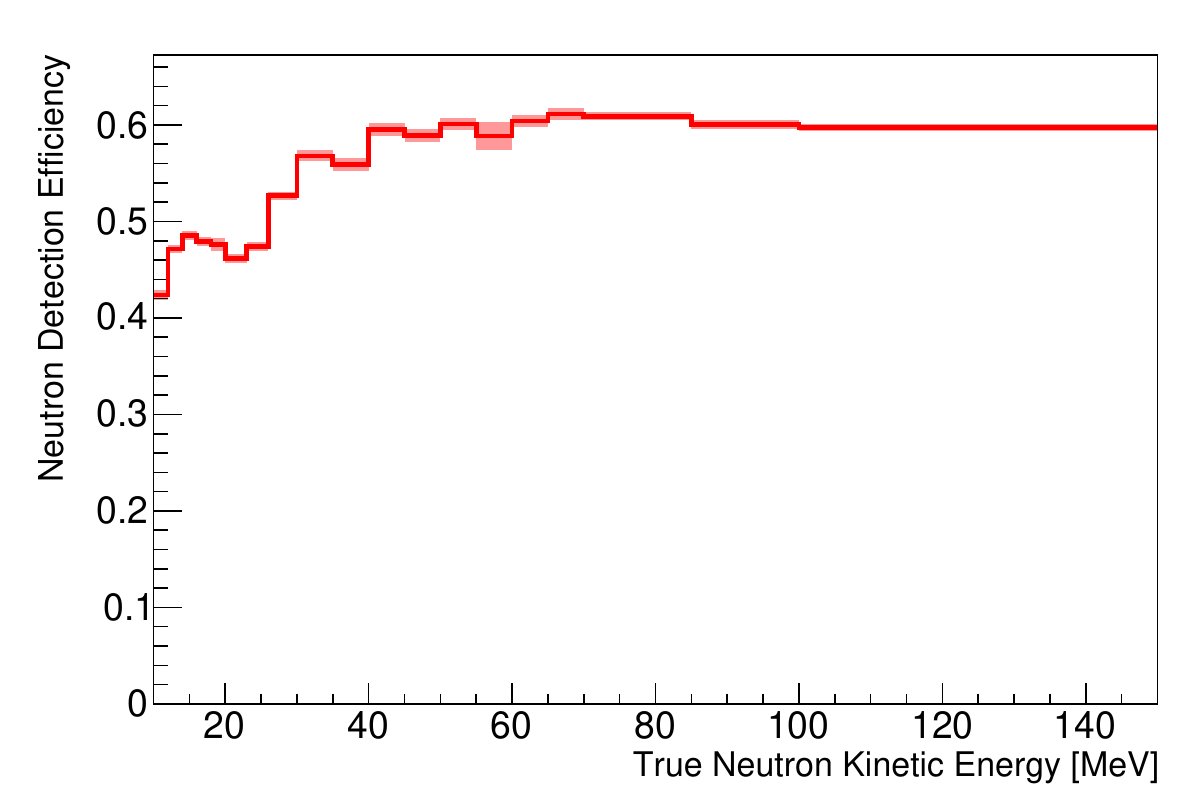}
  \caption{Efficiency to detect one or more neutron candidates from a final state neutron produced by an antineurtrino interaction in MINERvA's tracker as a function of neutron kinetic energy.  Antineutrino interactions had to be inside an 850~mm apothem hexagon at the center of MINERvA's tracker.  The simulated angular distribution of neutrons produced was driven by MnvTunev1 which is based on GENIE 2.12.6.  The pink error band includes systematic and statistical uncertainties.  Notice that neutrons below 10~MeV of kinetic energy are not included.}
  \label{fig:neutronDetectionEfficiency}
\end{figure}

MINERvA detects neutrons when they scatter inelastically.  Most neutron scattering interactions are visible in MINERvA data through either MeV-scale recoil protons or de-excitation photons from excited carbon nuclei.  Neutron reconstruction in MINERvA starts by searching for charge deposits that are isolated from the antineutrino interaction location, hereafter called the event vertex.  A key challenge for this algorithm is to avoid untracked activity from charged particles near the event vertex or the end of a particle's path through the detector.  Charge deposits close to the muon-hadron system are recursively excluded from the search for neutron activity.  First, all charge deposits that are part of the muon track are labelled as charged particle activity and excluded from the search for neutron activity.  Charge deposits that are likely to be cross-talk activity through adjacency to other electronics channels with high charge deposits or below a 1~MeV electronics noise threshold (low activity) are also excluded.  Next, any charge deposit within three strips of an excluded charge deposit is added to the list of excluded clusters.  This procedure repeats until no new charge deposits are excluded.  Figure \ref{fig:connectedClusterFilter} illustrates the typical end result of this procedure with charged hadron candidates.  Apparent charged hadron activity near the interaction vertex (purple) is left out of the search for neutron activity.  Neutron candidates directly transverse to and just behind the antineutrino interaction vertex are retained which improves neutron detection efficiency for the lowest energy neutrons over Ref. \cite{MirandaThesisPaper}.  When there is little charged hadron energy and neutrons are produced in the angular distribution predicted by MINREvA's antineutrino interaction simulation, this allows neutrons around 10-50~MeV to be identified with nearly 50\% efficiency as shown in Fig. \ref{fig:neutronDetectionEfficiency}.

Distinct neutron candidates are formed from the remaining hadron-like charge deposits using a similar algorithm to that used in Ref. \cite{TejinThesisPaper}.  First, a neutron candidate seed in two dimensions was formed from the most energetic grouping of charge deposits in each scintillator strip orientation (view).  Next, seeds that overlapped other seeds within the same view in transverse position were merged if they were close enough in longitudinal position.  Any lone charge deposits were promoted to two-dimensional seeds at this point.  Then, seeds were merged across views if they overlapped in longitudinal position and transverse position.  Seeds with charge deposits from two or more views became three-dimensional neutron candidates.  Remaining single-view seeds became two-dimensional neutron candidates.  Unlike Ref. \cite{TejinThesisPaper}, both types of neutron candidates were used in this analysis.  Any neutron candidates with less than 1.5~MeV deposited were not used for this result.  Combining spatially adjacent charge deposits and allowing two-dimensional neutron candidates in this way gives a more accurate count of the number of neutrons produced in the final state than the techniques used in either Ref. \cite{TejinThesisPaper} or Ref. \cite{MirandaThesisPaper}.  A single neutron could still scatter multiple times in MINERvA producing multiple neutron candidates.  This reconstruction algorithm makes no attempt to group scatters from the same neutron.

Reconstructed $E_{available}$ is defined as all non-muon energy that is not tagged as neutron activity.  It is designed to estimate energy transfer to the hadronic system with reduced dependence on energy deposited by neutrons which are hard to reconstruct.  Because most selected events have significant energy leaving the detector via neutrons, $E_{available}$ is only a rough proxy for energy transfer in this analysis.  Available energy used in this signal definition includes the total energy of neutral pions, all kaons, electrons, and photons and only the kinetic energy of protons and charged pions.  This has been found to better represent the total kinetic energy that was converted to charged hadron activity.

\section{Simulation}
\label{sec:Simulation}
The $\bar{\nu}_{\mu}$ energy distribution (flux) at MINERvA was simulated using PPFX \cite{PPFX} and GEANT4.  First, proton interactions on the NuMI beam's graphite target were simulated using GEANT4's \detokenize{FTFP_BERT} physics list \cite{GEANT4}.  Next, PPFX reweighted the predictions for hadrons leaving the target to match cross sections from MIPP \cite{MIPP}, NA49 \cite{NA49}, and some other smaller datasets.  Then, a GEANT4-based package was used to simulate these hadrons traveling through the beamline's focusing magnets and decaying into antineutrinos.

Antineutrino interactions in the MINERvA detector were simulated using GENIE 2.12.6 \cite{GENIE} \cite{GENIEApps}.  2p2h interactions were simulated using the Valencia model \cite{Valencia2p2h} \cite{Nieves2p2h} \cite{Schwehr2p2h}.  Quasielastic interactions were modeled using Llewellyn Smith's formalism with an axial mass of $M_A^{QE}$ = 0.99~$GeV/c^2$ \cite{LlewellynSmith}.  The Rein-Seghal model was used to simulate resonance production with an axial mass of $M_A^{RES}$ = 1.12~$GeV/c^2$ \cite{ReinSeghal}.  The nuclear medium was modeled as a relativistic Fermi gas (RFG) with a Bodek-Ritchie tail at high momentum \cite{BodekRitchie}.  Deep inelastic scattering (DIS) was simulated using the Bodek-Yang model \cite{BodekYang} with hadronization handled by the AGKY hadronization model \cite{AGKY}.  AGKY used KNO scaling \cite{KNOScaling} below invariant mass of W = 2.3~GeV and interpolated to PYTHIA 6.4 at W > 3.0~GeV \cite{Pythia}.  Final state interactions (FSI) were simulated using GENIE 2.12.6's hA simulation which is an effective single-step algorithm \cite{Intranuke} \cite{HarewoodFSIBugFix}.

The central value (CV) simulation is reweighted to MnvTunev1, an empirical tune to MINERvA low energy (LE) low recoil neutrino scattering data \cite{LELowRecoil}.  This tune is believed to be an appropriate starting point for MINERvA antineutrino data because it correctly predicted the LE antineutrino version \cite{LELowRecoilAntineutrino} of the calorimetric neutrino samples \cite{LELowRecoil} used to make the tune.  Notably, MnvTunev1 greatly enhances the rate of 2p2h interactions in the region between the peak rate of CCQE interactions and the peak rate of delta resonance production.  The CCQE process is modified using the long range correlations screening technique known as the random phase approximation, RPA, using a reweight that reproduces the predictions of the Valencia group \cite{ValenciaRPA} \cite{GranRPAUncertainty}.  Non-resonant pion production was reweighted to match constraints from re-analyzed bubble chamber data \cite{BubbleChamberReAnalysis}.  The magnitude of the 3-momentum transfer range of the Valencia model was extended to 2.0 GeV/c as described in Ref. \cite{MarvinThesisPaper}.

The extracted cross section will be compared to several other leading models in the field.  NUISANCE \cite{NUISANCE} was used to generate predictions for the low $E_{available}$ multi-neutron cross section using various configurations of GENIE v3 \cite{GENIE3CrossSectionTune} \cite{GENIE3FSITune}.  The GENIE 2.12.6 simulation was reweighted to approximate the SuSA v2 2p2h model without changing the CCQE model \cite{SuSA2p2h} \cite{SuSAInGENIE} as described in Ref. \cite{MarvinThesisPaper}.  The Valencia model compared to the data is the same 2p2h model used in GENIE 2.12.6 without the MnvTunev1 reweights.

Particle transport was simulated using GEANT 4.9.3 p02.  The neutron inelastic cross section was reweighted to more recent neutron interaction data as described in Ref. \cite{MirandaThesisPaper}.  

\section{Neutron Counting in 6 GeV MINERvA Data}
MINERvA first measured the rate of neutron-induced activity in the Low Energy (LE) data in Ref. \cite{MirandaThesisPaper}.  This first era of MINERvA data used an (anti)neutrino beam with average (anti)neutrino energy of 3 GeV.  Figure 5 of Ref. \cite{MirandaThesisPaper} showed that the LE-era simulation predicted a rate of neutron candidates that is significantly different from that which was observed in data.  Modifications to GENIE and GEANT were compared to the data, but neither modified simulation explained the data in all regions of phase space studied.  This result reproduces that study using Medium Energy-era (ME) data, which used a neutrino beam with average energy of 6 GeV at higher intensity.  The ME data has a much smaller statistical uncertainty,  and it extends across a broader phase space in neutrino kinematics.

\begin{figure}[bhtp]
  \includegraphics[scale=0.35]{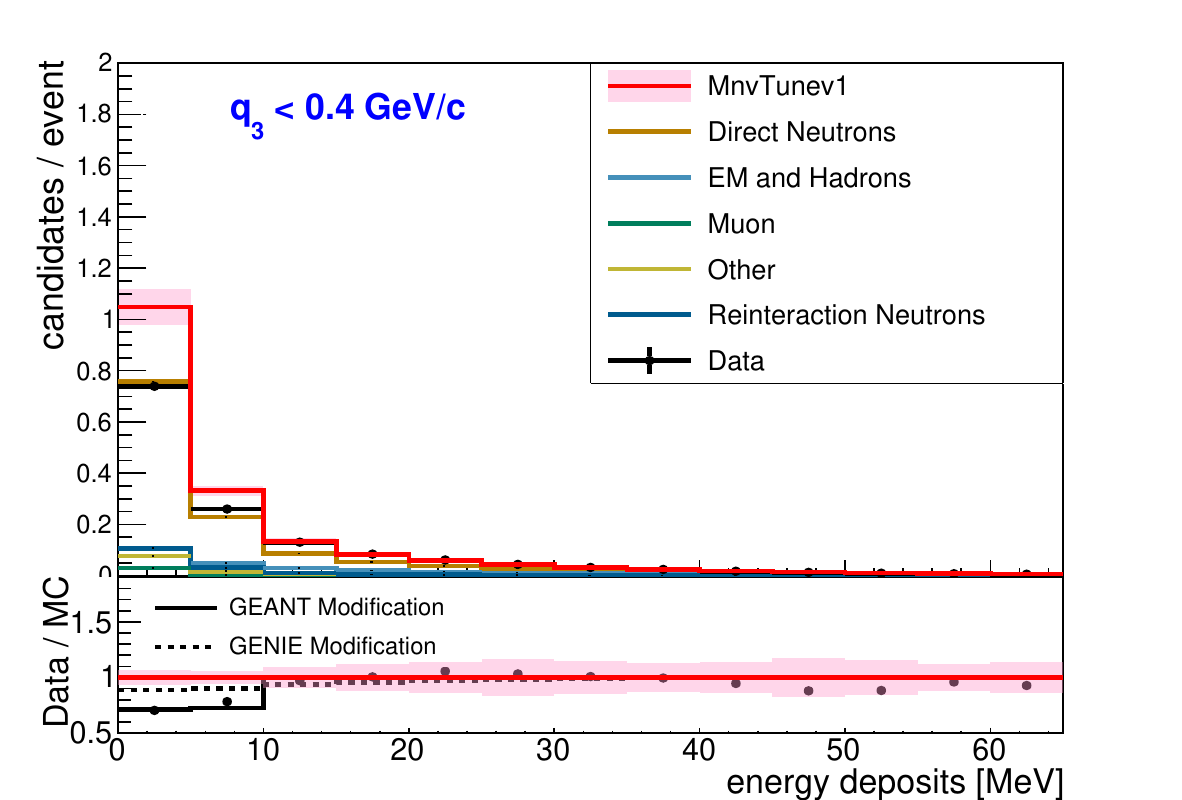}
  \includegraphics[scale=0.35]{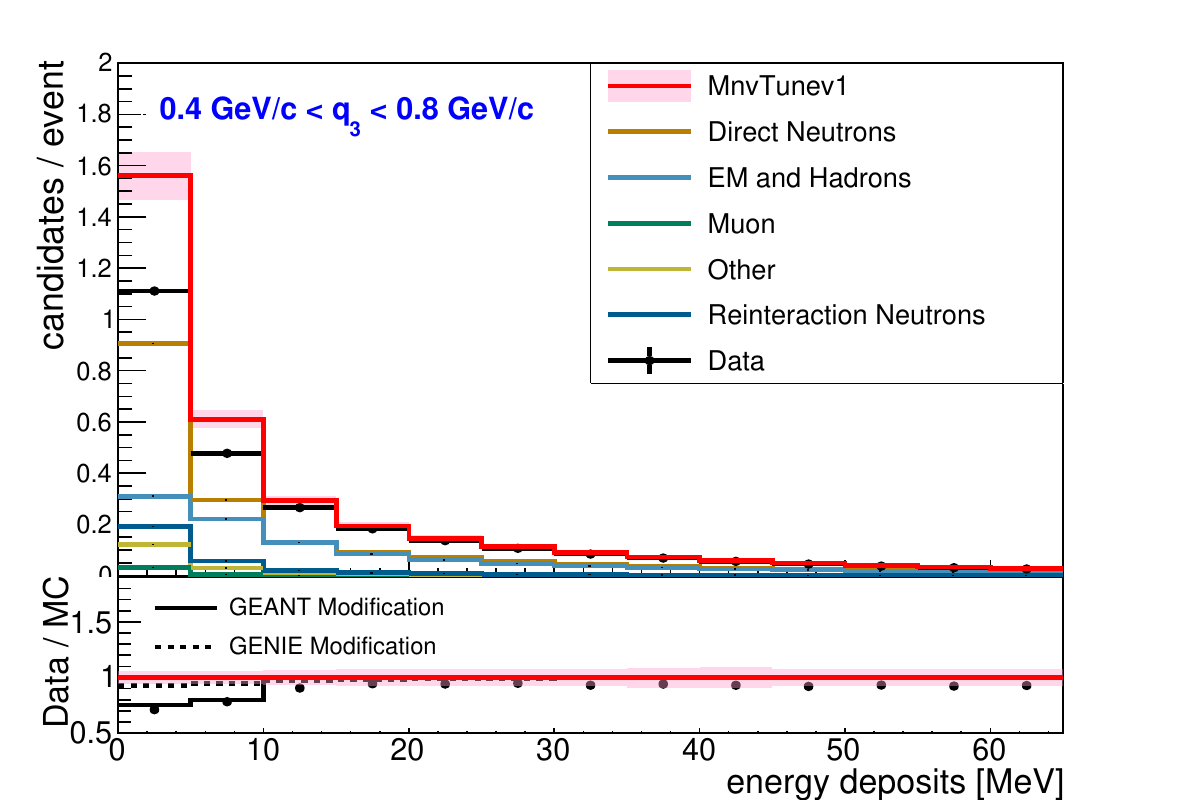}
  \caption{Neutron candidates per event in bins of energy deposited at $q_3$ < 0.4~GeV/c (top) and 0.4~GeV/c < $q_3$ < 0.8~GeV/c (bottom) where $q_3$ is the magnitude of the 3-momentum transfer.  In the top panels, MC candidate rate predictions are broken down into the type of particle that was reconstructed as a neutron candidate.  The bottom panels compare the data/MC ratio to two alternative models.  Error bands on the data points include only statistical uncertainties.  The pink uncertainty envelope on the MC lines carries all systematic uncertainties discussed in this paper.}
  \label{fig:candidateEDeps}
\end{figure}

Figure \ref{fig:candidateEDeps} shows the number of neutron candidates present in two regions of momentum transfer in the same format as Ref. \cite{MirandaThesisPaper}.  Reconstructed $q_{3}$ was calculated by using a spline to estimate total hadronic energy, treating this as energy transfer, and subtracting it from $Q^2$ as in Ref. \cite{MirandaThesisPaper}.  Figure \ref{fig:candidateEDeps} shows 0.75 neutron candidates per antineutrino interaction in the lowest candidate energy deposit bin which is higher than the 0.3 candidates per interaction in Ref. \cite{MirandaThesisPaper}.  This is a consequence of the more efficient neutron tagging algorithm developed for this result.  Most of the extra neutron candidates are located close enough to the interaction vertex that the algorithm in Ref. \cite{MirandaThesisPaper} would have excluded them.  A similar number of candidates per event can be obtained in ME-era data by using the LE-era algorithm.

The lower panel of Fig. \ref{fig:candidateEDeps} highlights the difference between ME-era neutron tagging data and simulation.  The Monte Carlo over-predicts the rate of neutron candidates with less than 10~MeV energy deposited.  This over-prediction does not extend to higher energy deposits.  Alternative models for antineutrino neutron production and neutron transport were tested as was done in Ref. \cite{MirandaThesisPaper}.  The GEANT modification drops 35\% of neutron candidates with energy deposit less than 10~MeV.  This modification models a generic change that could be ascribed to GEANT's de-excitation model for example and could include shortcomings in neutron modelling like the neutron inelastic cross section mis-modeling discussed in Ref. \cite{MoNAPaper}.  The GENIE modification drops half of all candidates that are produced by final state (FS) neutrons with less than 50~MeV of kinetic energy.  This generic change might mimic a major shortcoming of the FSI model or a change in the strength of the 2p2h process.  MINERvA observes the same discrepancy in modeling the lowest energy deposit neutron candidates across both this ME-era result and the LE-era result in Ref. \cite{MirandaThesisPaper}.

\section{Cross Section Extraction}
A single-differential cross section was extracted as a function of muon transverse momentum, $p_{T \mu}$, for a $\bar{\nu}_{\mu}$ to produce two or more neutrons when $E_{available}$ is less than 100~MeV in a charged-current interaction.  Each neutron in the signal definition was required to have at least 10~MeV of kinetic energy because MINERvA loses most of its sensitivity to neutrons below this threshold.  Limiting $E_{available}$ to less than 100~MeV preferentially selects 2p2h interactions and gets rid of interactions with neutral pions and high energy final state charged hadrons that tend to produce additional neutron-like activity.  Muon $p_{T}$ is interesting because it provides insight into four-momentum transfer similar to $Q^2$ in this kinematic region without introducing additional dependence on neutron modeling \cite{LEAntineutrinoQE}.  It is also easier for models to predict than hadronic energy estimators for MINERvA \cite{LENeutrinoQE}, especially when multiple neutrons are produced.  Searching for events with at least two neutrons enhances the proportion of 2p2h interactions selected because the presence of the second neutron greatly reduces the contribution from CCQE interactions.  Allowing more than two neutrons reduces the dependence of the cross section on neutron production and interaction modeling.

Equation \ref{eq:crossSectionFormula} shows the components used to extract the cross section from the data.

\begin{equation} \label{eq:crossSectionFormula}
  \frac{d\sigma_{signal}}{dp_{T \mu}}_{j} = \frac{\Sigma_i U_{ij}(N^{sel}_i - \Sigma_k \alpha_{ik} \times N^{sel}_{bkg, ik})}{\epsilon_j \Phi \Delta p_{T \mu j} N_{nucleons}}
\end{equation}

Here, $j$ is a bin number in true $p_{T \mu}$, and $i$ is a bin number in reconstructed $p_{T \mu}$.  $U$ is an unsmearing function described later in this section.  $N_{i}^{sel}$ is the number of data interaction candidates selected.  $\alpha_{ik}$ is a background scale factor.  $N_{bkg,ik}^{sel}$ is the number of background MC interactions selected.  $\epsilon_{j}$ is the product of efficiency and acceptance for the signal definition estimated using the simulation of MINERvA.  $\Phi$ is the flux integrated across all energy bins.  $\Delta x_j$ is the width of bin j.  $N_{nucleons}$ is the number of nucleons in the fiducial volume of the MINERvA detector.

Selected events must have $E_{available}$ less than 100~MeV as defined above.  They must have at least two neutron candidates that are within 1.5~m of the interaction point and each have at least 1.5~MeV of energy deposited.  Neutron candidates more than 1.5~m from the interaction vertex are much more likely to come from uncorrelated backgrounds than from the antineutrino interaction being studied.  The 1.5~MeV energy deposit requirement cuts out the vast majority of cross-talk activity \cite{MirandaThesisPaper}.  Antineutrino interactions that produce muons at angles of greater than 20~degrees relative to the beam direction are not counted because they are not efficiently reconstructed by MINOS.  Muons must have at least 2~GeV/c of momentum to reliably appear in MINOS and less than 20~GeV/c of momentum to originate from the region of best-understood flux.  Only antineutrino interactions within an 850~mm apothem hexagon of the center of the active tracker were used to extract this cross section.  After all cuts, the selected sample is estimated to have 39\% efficiency and 39\% purity with 299182 interactions selected in data.

\begin{figure}[ht]
  \includegraphics[scale=0.4]{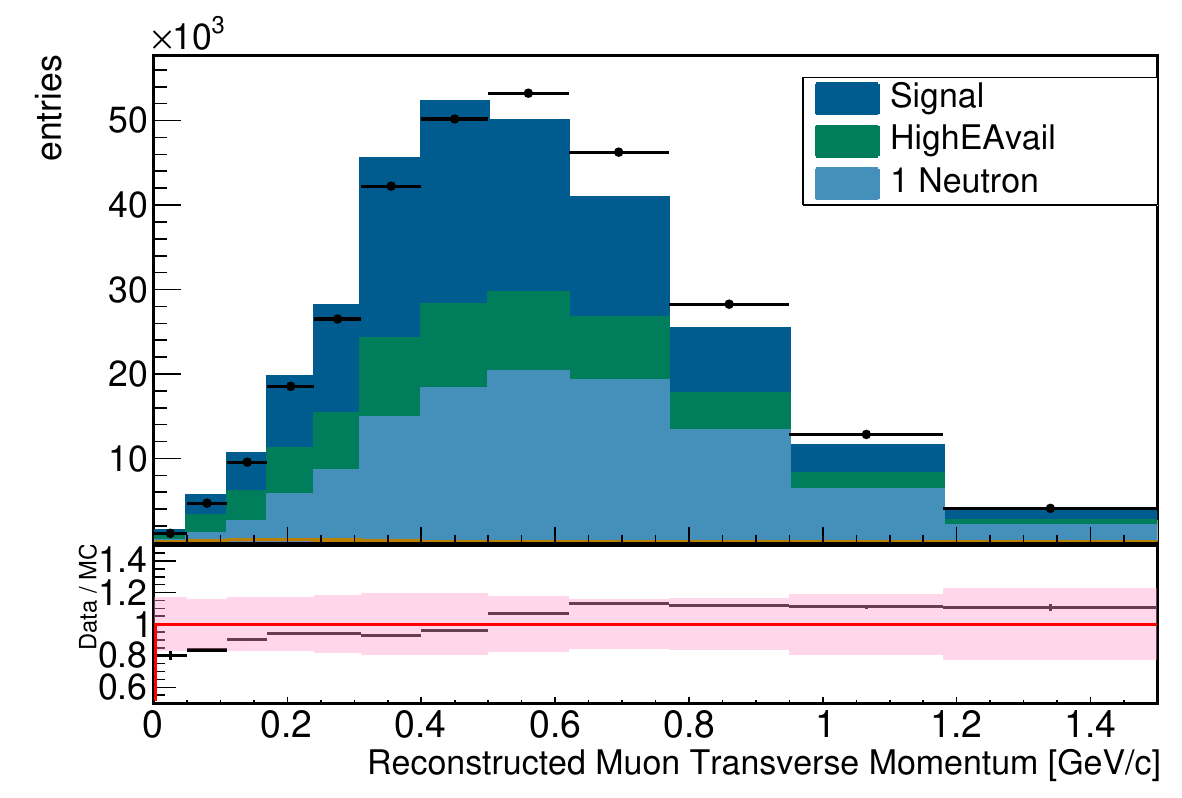}
  \includegraphics[scale=0.4]{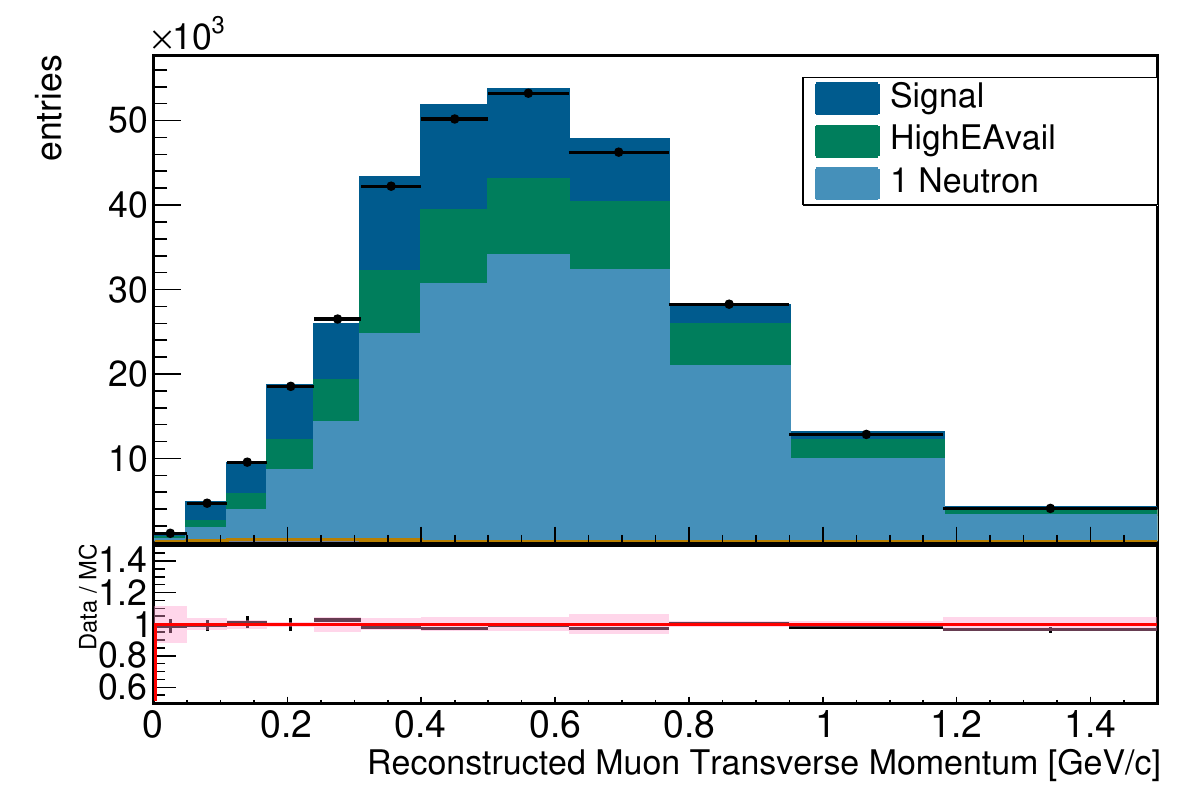}
  \caption{Selected events in data and MnvTunev1 Monte Carlo without any background rate correction (left) and with background rate correction (right).  Notice that the background correction matches the shape of the overall MC prediction better to data and reduces systematic uncertainty on the MC prediction (pink band in bottom panel).  Other backgrounds are less than 1\%.}
  \label{fig:selectionBeforeFit}
\end{figure}

Two leading backgrounds were identified: one background with one neutron in the final state and a another background where $E_{available}$ was mis-reconstructed.  The one neutron backgrounds are predicted to be mostly CCQE antineutrino interactions.  A muon antineutrino CCQE interaction produces a positive muon and a single neutron before FSI.  The neutron could produce multiple neutron candidates by interacting multiple times in the MINERvA detector.  The high $E_{available}$ background comes mostly from interactions where a pion was produced in the final state and reconstructed as a neutron candidate.  Neutral pions are an obvious background to neutron tagging.  In this analysis they are usually excluded by the $E_{available}$ cut.  Low energy charged pions can be reconstructed as neutron candidates if they travel transverse to the beam direction.  Figure \ref{fig:selectionBeforeFit} shows the $p_{T \mu}$ distribution of backgrounds estimated by MINERvA simulation.

\begin{figure}[ht]
  \includegraphics[scale=0.4]{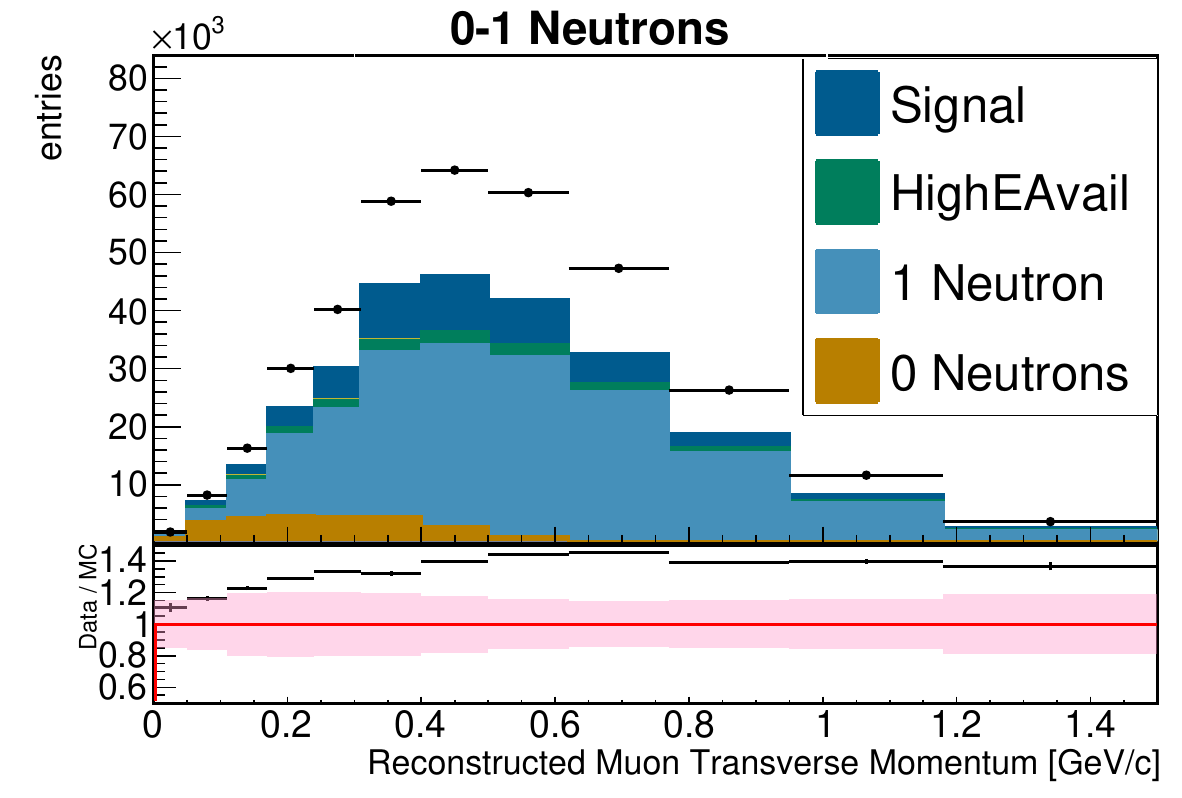}
  \includegraphics[scale=0.4]{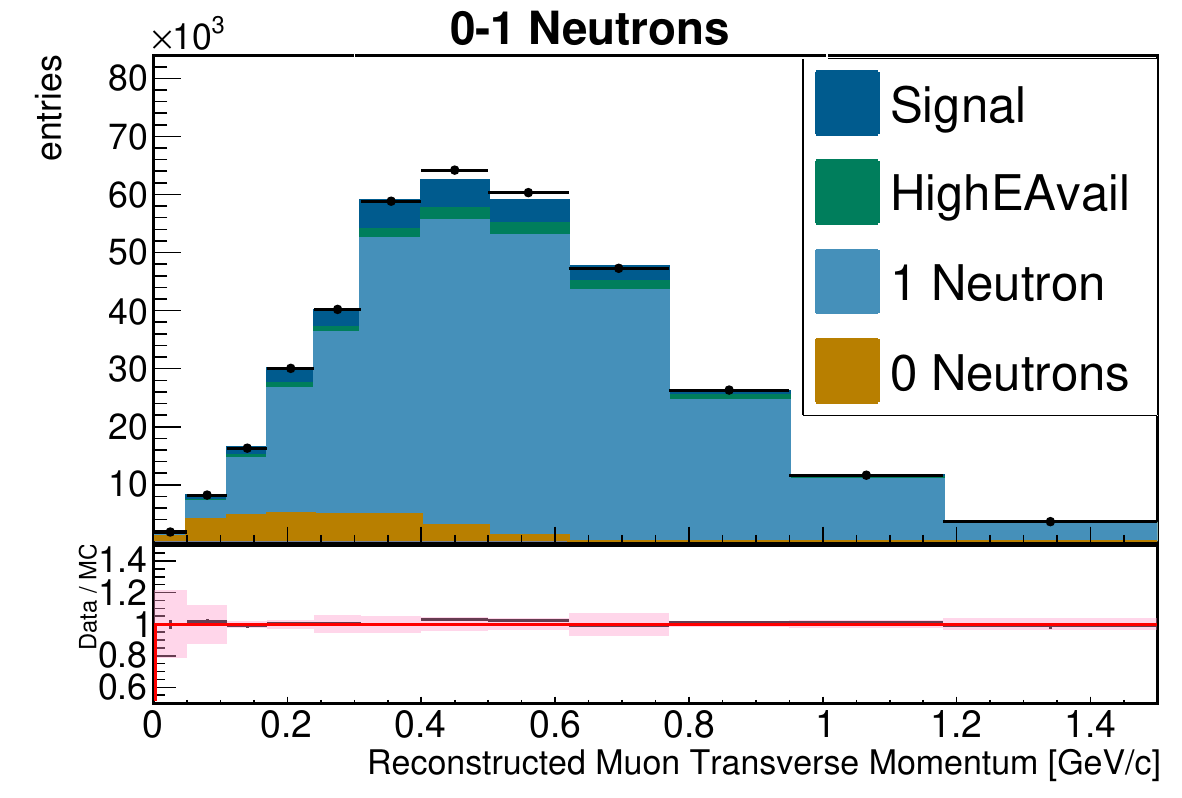}
  \caption{Few neutrons selection in data and MnvTunev1 Monte Carlo without any background rate correction (left) and with background rate correction (right).  The Monte Carlo model predicts this selection to be very rich in the 1 neutron background.  The background correction improves shape agreement with the data and reduces systematic uncertainty.  This is the only sample where the 0 neutrons background contributes appreciably.}
  \label{fig:fewNeutronsBeforeFit}
\end{figure}

\begin{figure}[ht]
  \includegraphics[scale=0.4]{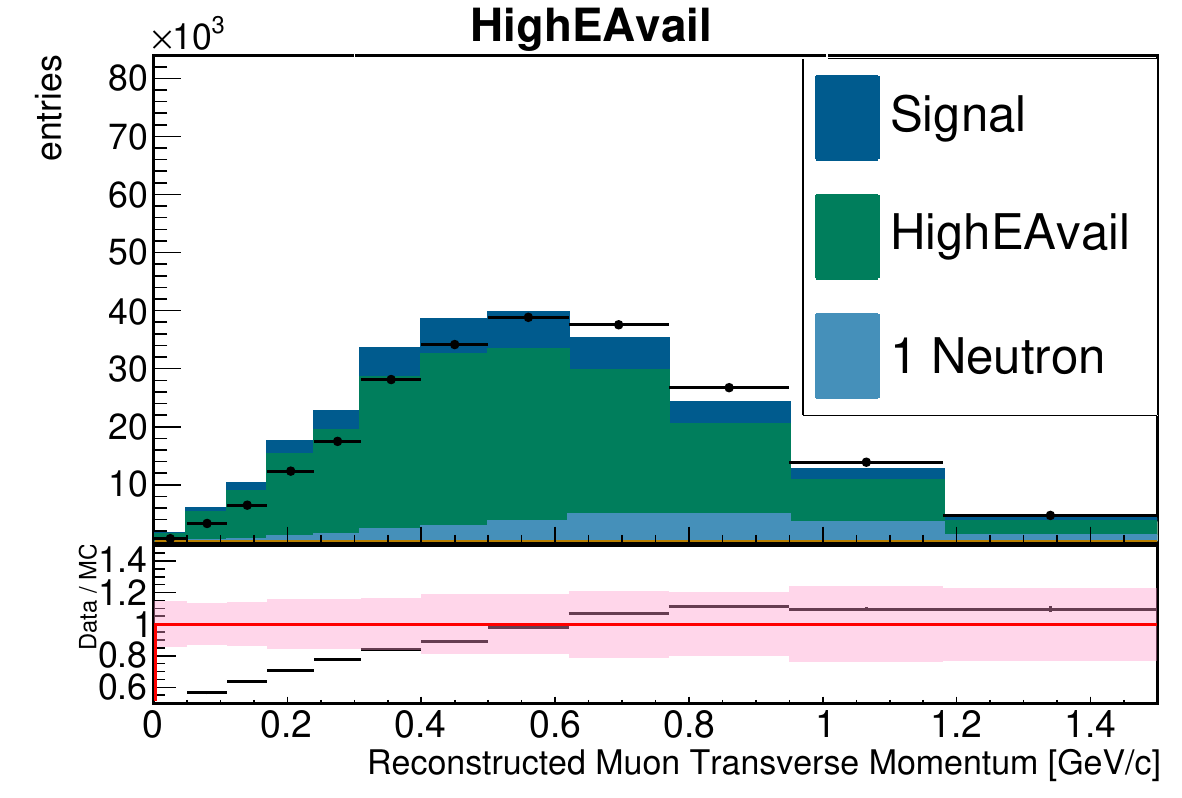}
  \includegraphics[scale=0.4]{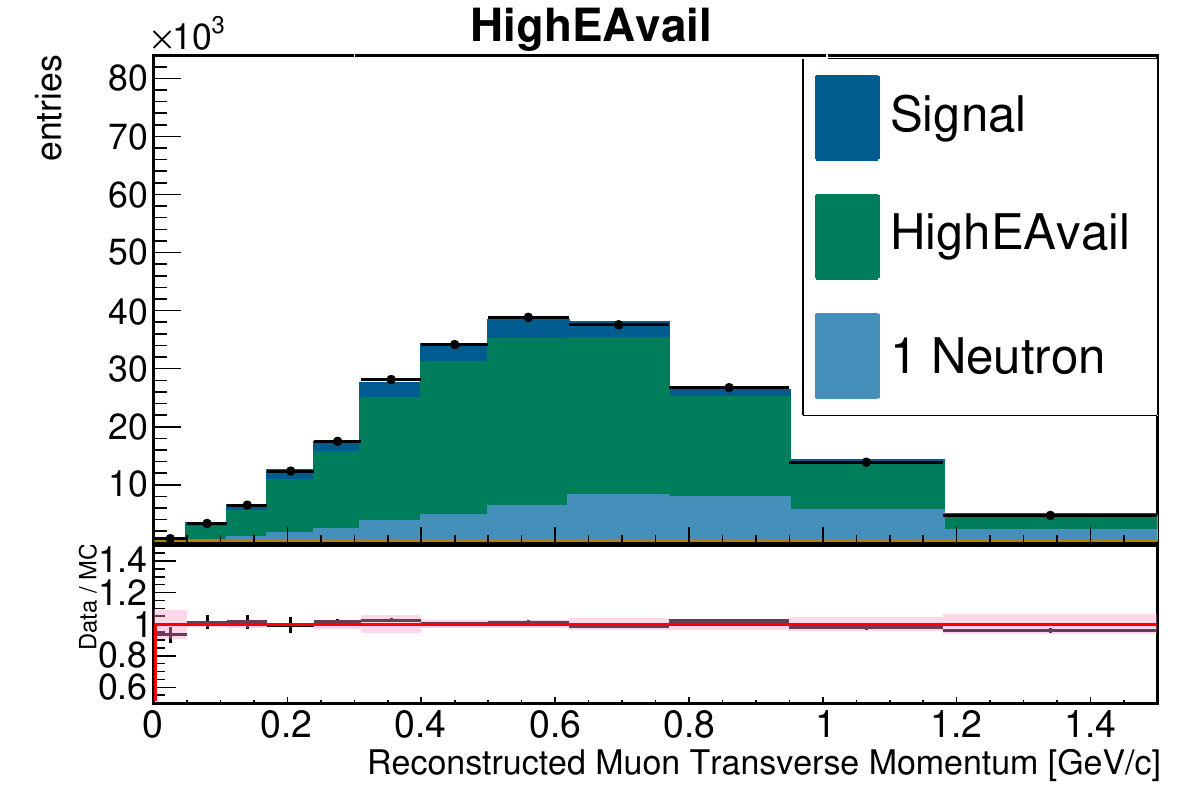}
  \caption{High $E_{available}$ selection in data and MnvTunev1 Monte Carlo without any background rate correction (left) and with background rate correction (right).  The Monte Carlo model predicts this selection to be rich in the high $E_{available}$ background.  The background correction improves shape agreement with the data and reduces systematic uncertainty as in the selection region and the other sideband region.}
  \label{fig:highEAvailBeforeFit}
\end{figure}

\begin{figure}[bhtp]
  \centering
  \includegraphics[scale=0.4]{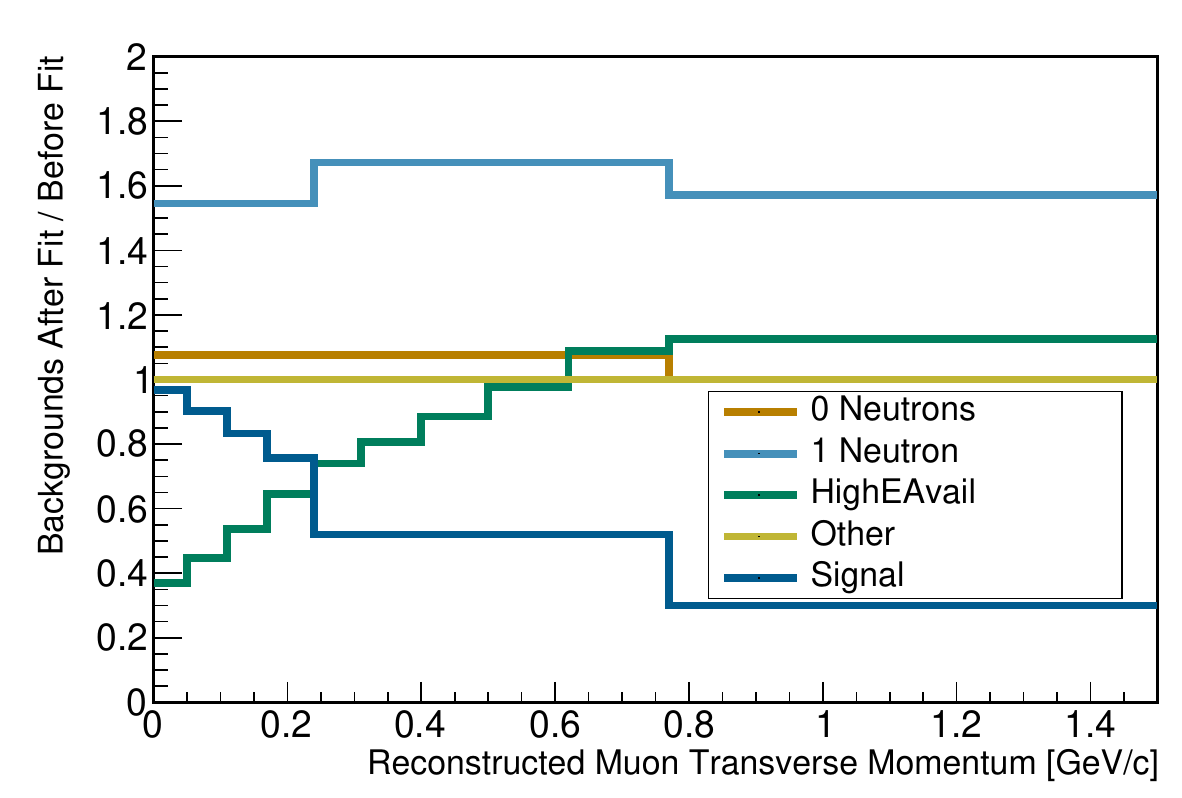}
  \caption{Summary of sideband fit parameters obtained by jointly fitting two sidebands and the selection region.  Notice how much the signal has been scaled down to find the best match to the data.}
  \label{fig:sidebandFitParameters}
\end{figure}

Backgrounds were constrained by data in sidebands to  reduce the model dependence of the extracted cross section.  Two sideband regions were selected: the one neutron sideband that is rich in single neutron events, and the 100~MeV < $E_{available}$ < 200~MeV sideband that is rich in events with high $E_{available}$.  Each sideband passes all other selection cuts so that it is similar to the selection region.  A joint fit was performed to each background distribution across all sidebands and the selection region.  The one neutron background was allowed to scale in 3 regions of $p_{T \mu}$ independently, and the high $E_{available}$ background was fit with two linear functions, from 0~MeV/c to 310~MeV/c and from 310~MeV/c to 770~MeV/c, and a scale factor above 770~MeV/c.  The signal model was also fit with a linear function in the lowest $p_{T \mu}$ region and scale factors in the other two regions.  Figure \ref{fig:sidebandFitParameters} shows the scaling from the fits applied as a function of muon transverse momentum.  The one neutron background was scaled up by 60\% which is consistent with the ME-era MINERvA $\bar{\nu}_{\mu}$ CCQE-like cross section measurement \cite{AmitsThesisPaper}.  The suppression of the high $E_{available}$ background looks similar to the low $Q^2$ pion suppression that was indicated in a joint fit to MINERvA LE pion data \cite{lowQ2PionSuppression} \cite{MINOSLowQ2}.  The signal scaling was only used to better estimate the background fits in the sideband regions and was not used when estimating efficiency or unsmearing the cross section.

Detector smearing in $p_{T \mu}$ was corrected using D'Agostini's unfolding procedure \cite{dAgostini} \cite{dAgostiniErrors}.  The simulation was used to estimate a smearing matrix.  This matrix was updated in steps prescribed by D'Agostini's procedure.  A large number of update steps, or iterations, reduces the bias in the final result at the cost of greater variance.  This result uses the number of iterations that minimizes bias without variance diverging.  The degree of bias at each iteration was estimated by unfolding selected distributions from many different models using the final unsmearing matrix.  Changing the 2p2h model of the antineutrino scattering simulation tended to produce different unfolded distributions for different models.  The D'Agostini unfolding procedure was truncated at three iterations where the differences between 2p2h models are smallest while the variance has not yet increased much.  An additional uncertainty covers the difference between unfolded results for the two most different 2p2h models as in Ref. \cite{MarvinThesisPaper}.  The unsmearing procedure was also checked for stability under statistical fluctuations.

The MnvTunev1 simulation was used to estimate the efficiency and acceptance of this selection.  The efficiency decreases gradually with $p_{T \mu}$.  Little uncertainty enters through this correction compared to the background correction.  The flux integral used to extract this cross section result was constrained by MINERvA neutrino- and antineutrino-dominated $\nu$-e elastic scattering and inverse muon decay constraints \cite{NuEConstraintPaper}.  The data used in this result corresponds to $11.1 \times 10^{21}$ protons on target (POT).

\section{Results}
\label{sec:Results}
\begin{figure}[bhtp]
  \centering
  \includegraphics[scale=0.4]{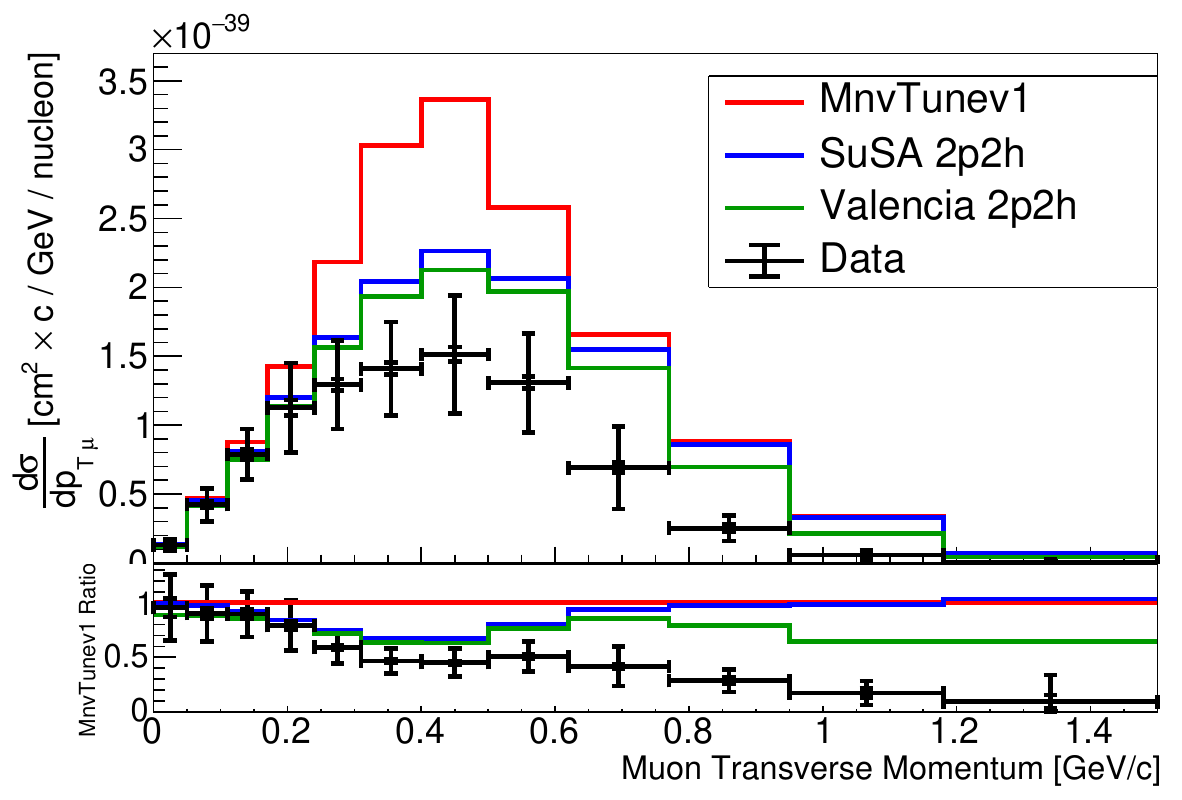}
  \caption{Cross section for an antineutrino to produce multiple neutrons in the final state and no more than 100~MeV of available energy.  The black data points are the result extracted from data using MnvTunev1 MC.  Inner error bars show statistical uncertainty, and outer error bars show the full uncertainty on each data point.  The colored lines are cross section predictions from various MC models available as reweights of MINERvA MC.  The bottom panel shows the ratio of each cross section to MnvTunev1.}
  \label{fig:crossSectionMnvTune}
\end{figure}

Figure \ref{fig:crossSectionMnvTune} presents the $\bar{\nu}_{\mu}$ multi-neutron production cross section at low $E_{available}$.  The uncertainties on the measurement are smaller than the differences between leading models.  Notice that the MnvTunev1 cross section estimate from the simulation is about 50\% higher than the data at its peak.  MnvTunev1 enhances the rate of 2p2h interactions by as much as 100\% over the Valencia model in many regions of the phase space studied.  Using the unmodified Valencia model or the SuSA v2 2p2h model produces predictions much closer to the data.  These models' predictions are still outside the data uncertainties at their peaks.

The bottom panel of Fig. \ref{fig:crossSectionMnvTune} shows the ratio of the cross section measured in the data and that predicted by each model to the cross section predicted by MnvTunev1.  The data agrees well with MnvTunev1 and most models at very low $p_{T \mu}$.  MnvTunev1 over-estimates the data rate where the cross section peaks.  The Valencia and SuSA models predict rates closer to the data rate than MnvTunev1.  At $p_{T \mu}$ > 0.6~GeV/c, the data cross section drops off precipitously.  As shown in Fig. \ref{fig:processBreakdown}, 2p2h is no longer predicted to be the dominant contribution to the signal in this region of phase space.  The Valencia model exhibits a slight downturn at high $p_{T \mu}$, but it is not nearly as strong as in the data.

\begin{figure}[bhtp]
  \centering
  \includegraphics[scale=0.4]{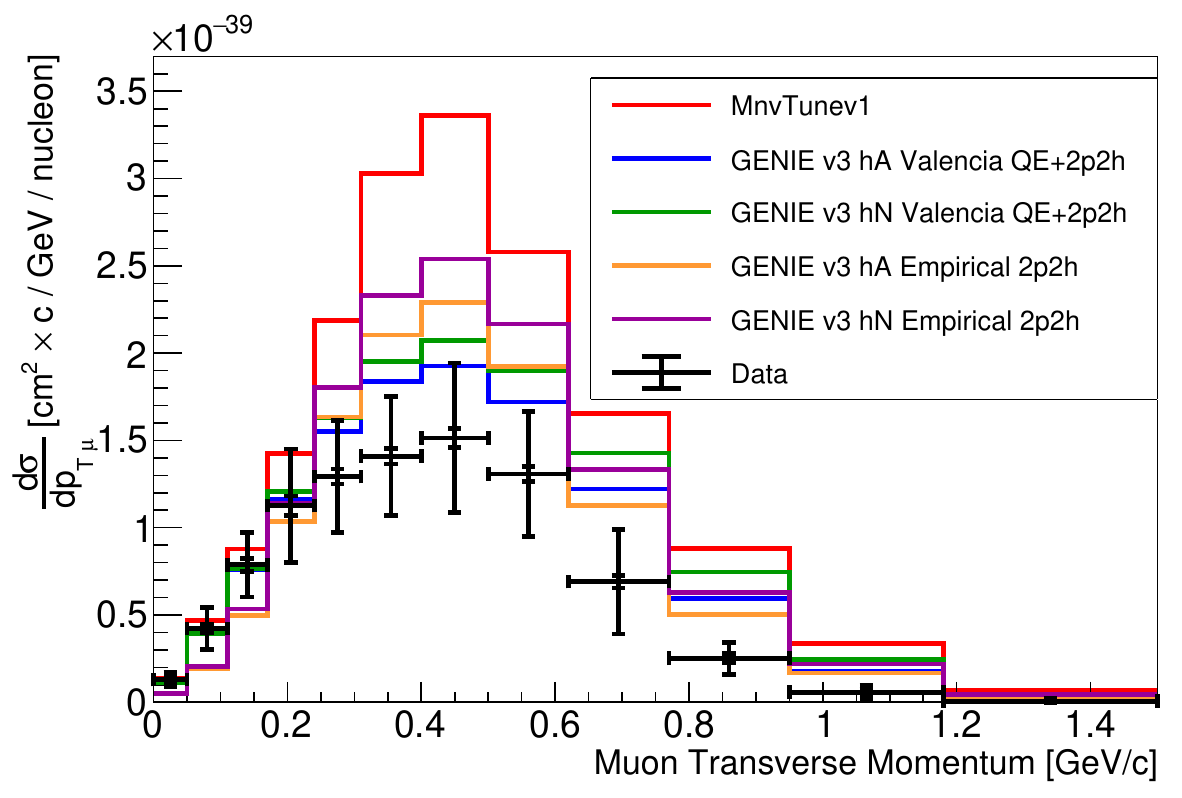}
  \caption{The same data-extracted cross section as Fig. \ref{fig:crossSectionMnvTune} now compared to various GENIE v3 configurations.  Empirical 2p2h is a tune in the same spirit as MnvTunev1 performed by Dytman \cite{GENIE}.}
  \label{fig:crossSectionGENIE}
\end{figure}

Figure \ref{fig:crossSectionGENIE} compares the data to models from GENIE v3 with different 2p2h and FSI simulations.  The GENIE v3 models all have a local Fermi gas CCQE model \cite{ValenciaRPA} instead of the Bodek-Ritchie tail, and they all use the Berger-Sehgal pion model \cite{Bercellie} with different form factors for resonance production instead of the Rein-Seghal model in MnvTunev1.  These GENIE v3 models differ most from each other in their treatment of FSI and their 2p2h models.  The GENIE v3 empirical 2p2h model was developed with electron scattering data and a procedure much the same as the MnvTunev1 2p2h enhancement: add 2p2h events to the region between CCQE and resonance production until the model matches the data \cite{2p2hElectronScattering}.  Ref. \cite{MarvinThesisPaper} points out that there are many ways additional resonance or CCQE interactions could fill this region calorimetrically while making different predictions for the final state hadron content.

\begin{figure}[bhtp]
  \centering
  \includegraphics[scale=0.4]{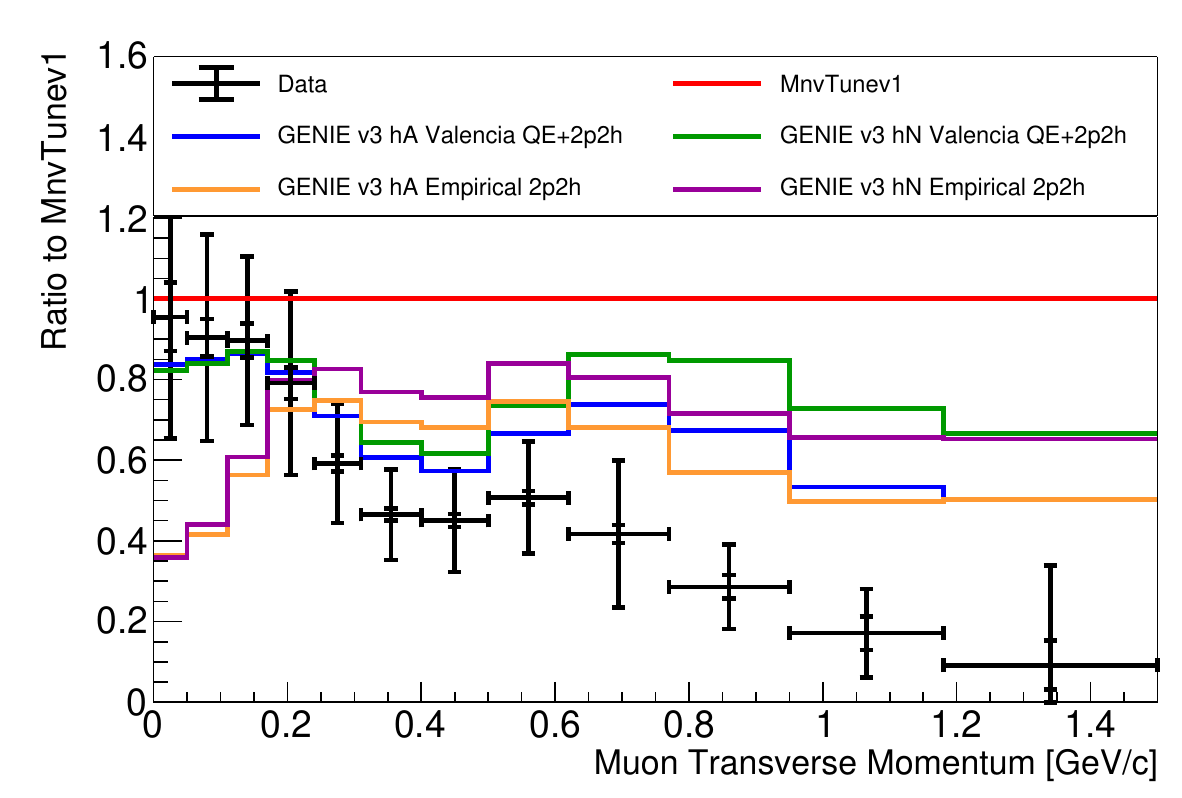}
  \caption{Ratio of the data-extracted cross section from Fig. \ref{fig:crossSectionMnvTune} to MnvTunev1.  This version compares the cross section to predictions from various configurations of GENIE v3.}
  \label{fig:crossSectionRatioGENIE}
\end{figure}

These GENIE v3 models all make predictions closer to the data than MnvTunev1 near the peak of the data distribution.  Valencia-based 2p2h models in GENIE v3, nearly identical to the Valencia model discussed earlier, make predictions of the peak region rate closer to the data than the empirical 2p2h models.  The hA FSI model matches the data rate better in the peak region than the more complex hN model.  Figure \ref{fig:crossSectionRatioGENIE} shows the ratio of each model to MnvTunev1 where again no model has a drop off as steep at high $p_{T \mu}$ as the data.

\begin{figure}[bhtp]
  \centering
  \includegraphics[scale=0.4]{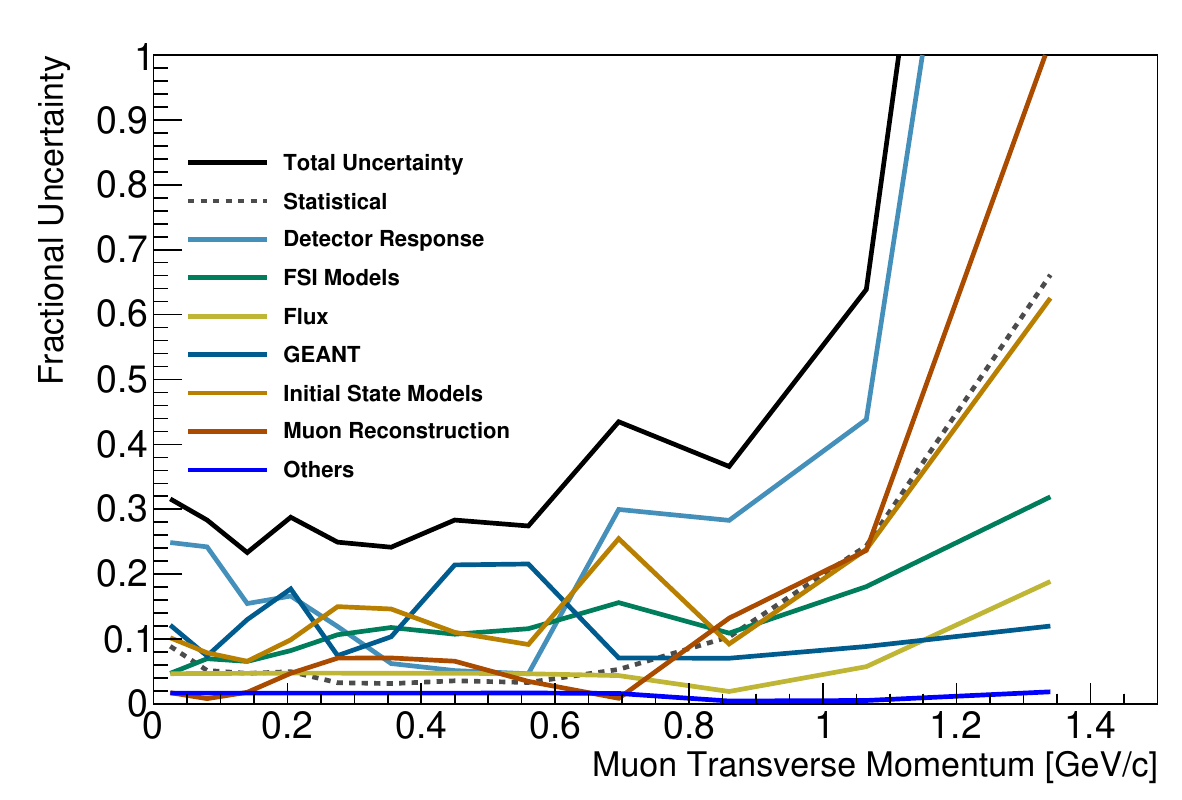}
  \caption{Breakdown of uncertainty sources in Fig. \ref{fig:crossSectionMnvTune}.  Each color represents a group of the error bands described in Section \ref{sec:Results}.  The black line shows the overall uncertainty as a fraction of the measured cross section.  It is the ratio of the size of the bars in Fig. \ref{fig:crossSectionMnvTune} to their respective data points.}
  \label{fig:crossSectionUncertaintySummary}
\end{figure}

Figure \ref{fig:crossSectionUncertaintySummary} summarizes how modeling assumptions contribute to the uncertainties in Fig. \ref{fig:crossSectionMnvTune}.  Statistical uncertainty is one of the smallest sources of uncertainty at about 6\% on average.  MINERvA typically estimates systematic uncertainties using the multi-universe method \cite{MATPaper}.  The systematic uncertainty is dominated across most of the $p_T$ range by neutron modeling in the GEANT band and 2p2h modeling in the Initial State Models band.  The Initial State Models band includes two 2p2h model assumptions: the MnvTunev1 2p2h enchancement and unfolding using the SuSA v2 2p2h model.  The MnvTunev1 2p2h model uncertainty is constructed from the RMS of enhancing only np or pp/nn nucleon pair interactions.  To cover any unfolding bias from the 2p2h model used, a cross section was extracted using MnvTunev1 reweighted with SuSA v2 as the MC simulation.  The uncertainty applied is half the difference between the CV and the SuSA v2 prediction which makes the full uncertainty cover the difference between unfolding models.

The "GEANT" error band is dominated by a neutron modeling uncertainty derived from an alternative neutron transport package.  Ref. \cite{MoNAPaper} explains how the MoNA collaboration deployed a scintillator detector made of the same material as MINERvA in a neutron test beam.  The data from this experiment was compared to both GEANT 4.9, the same simulation MINERvA uses, and an alternative neutron transport simulation called \detokenize{MENATE_R}.  The neutron detection rate in the MoNA detector matches the prediction from \detokenize{MENATE_R} much more closely than the prediction from GEANT.  \detokenize{MENATE_R} simulates MeV-scale neutron transport in greater detail than GEANT by using collected nuclear physics data to tune a separate cross section for each way a neutron can inelastically scatter off of a carbon nucleus.  For example, \detokenize{MENATE_R} has one data-driven cross section for a neutron to interact on a carbon nucleus and produce a proton and a neutron.  GEANT 4.9 is tuned to total inelastic cross section data for MeV-scale neutron interactions and relies on a cascade simulation to break down neutron inelastic interactions in detectors like MINERvA.  To calculate the "GEANT" error band in this cross section result, the CV model was reweighted to look as similar to the cross section data used in \detokenize{MENATE_R} as possible for each neutron inelastic cross section using a similar technique as was used for the overall neutron transport model described in Sec. \ref{sec:Simulation} of this paper.  Then, the difference between the reweighted model and the CV model was taken as an uncertainty.

\section{Conclusions}
MINERvA has measured neutrons per event in its 6~GeV data and sees a discrepancy that is consistent with its 3~GeV data from Ref. \cite{MirandaThesisPaper}.  Neither the GEANT modification nor the GENIE modification from Ref. \cite{MirandaThesisPaper} offers a clear explanation for what could be missing in MnvTunev1.

A cross section for multi-neutron production was extracted to test modeling of antineutrino neutron production on polystyrene in a model- and detector-independent manner.  MnvTunev1, which was designed to adapt to 2p2h-sensitive data, greatly overpredicts this cross section.  Other leading models get closer to matching the data, but no model matches the data rate well in the region where the data rate peaks.  Low $p_{T \mu}$ is described well by most models while high $p_{T \mu}$ has a much steeper drop off than any model predicts.  The extracted cross section is particularly interesting because of its sensitivity to 2p2h interactions and FSI.  MnvTunev1's enhancement of 2p2h does not seem to be helping it explain this data.  Other recent MINERvA cross section measurements \cite{MarvinThesisPaper} \cite{AmitsThesisPaper} exhibit similar behavior.  In the high $p_{T \mu}$ region, where 2p2h contributions are expected to be small, FSI mis-modeling could be driving the growing difference between data and simulation.  Other (anti)neutrino detectors, especially plastic-based detectors, should regard antineutrino 2p2h simulations with skepticism and consider evaluating a variety of neutron transport models.

\begin{acknowledgments}

This document was prepared by members of the MINERvA Collaboration using the resources of the Fermi National Accelerator Laboratory (Fermilab), a U.S. Department of Energy, Office of Science, HEP User Facility. Fermilab is managed by Fermi Research Alliance, LLC (FRA), acting under Contract No. DE-AC02-07CH11359.
These resources included support for the MINERvA construction project, and support
for construction also
was granted by the United States National Science Foundation under
Award No. PHY-0619727 and by the University of Rochester. Support for
participating scientists was provided by NSF and DOE (USA); by CAPES
and CNPq (Brazil); by CoNaCyT (Mexico); by ANID PIA / APOYO AFB180002, CONICYT PIA ACT1413, and Fondecyt 3170845 and 11130133 (Chile); 
by CONCYTEC (Consejo Nacional de Ciencia, Tecnolog\'ia e Innovaci\'on Tecnol\'ogica), DGI-PUCP (Direcci\'on de Gesti\'on de la Investigaci\'on  - Pontificia Universidad Cat\'olica del Peru), and VRI-UNI (Vice-Rectorate for Research of National University of Engineering) (Peru); NCN Opus Grant No. 2016/21/B/ST2/01092 (Poland); by Science and Technology Facilities Council (UK); by EU Horizon 2020 Marie Skłodowska-Curie Action; by a Cottrell Postdoctoral Fellowship from the Research Corporation for Scientific Advancement; by an Imperial College London President's PhD Scholarship.  We thank the MINOS Collaboration for use of its near detector data. Finally, we thank the staff of
Fermilab for support of the beam line, the detector, and computing infrastructure.

%
%
%
%
%
%
%
%
%
%

\end{acknowledgments}

\bibliography{MultiNeutronPaper.bib}
\bibliographystyle{unsrt}

\appendix
\section{Data Release}
\begin{table*}[tp]
 \begin{filecontents*}{crossSection.csv}
Bin Low Edge [GeV/c],0.00,0.05,0.11,0.17,0.24,0.31,0.40,0.50,0.62,0.77,0.95,1.18
Cross Section,0.1306,0.4228,0.7874,1.1262,1.2934,1.4095,1.5142,1.3077,0.6907,0.2521,0.0575,0.0062
Total Uncertainty,0.0413,0.1200,0.1844,0.3237,0.3264,0.4456,0.5479,0.3600,0.3768,0.0922,0.0370,0.0168
\end{filecontents*}
\csvautotabular{crossSection.csv}
 \caption{Bin edges, values, and uncertainties from the data cross section histogram in Fig. \ref{fig:crossSectionMnvTune}.  Cross section and uncertainty values in this table are to be scaled by  $10^{-39}$~$cm^{2} \times c / GeV / nucleon$.  The last bin ends at 1.5~GeV/c.}
\end{table*}


\begin{table*}[tp]
 \begin{filecontents*}{crossSectionCovariance.csv}
Bin Number,1,2,3,4,5,6,7,8,9,10,11,12
1,0.0017,0.0046,0.0061,0.0092,0.0052,-0.0011,-0.0034,-0.0006,0.0079,0.0020,0.0008,-0.0001
2,0.0046,0.0144,0.0187,0.0282,0.0194,0.0020,-0.0037,0.0048,0.0269,0.0068,0.0022,-0.0004
3,0.0061,0.0187,0.0340,0.0547,0.0389,0.0016,-0.0229,-0.0096,0.0288,0.0062,0.0031,-0.0008
4,0.0092,0.0282,0.0547,0.1048,0.0860,0.0247,-0.0277,-0.0279,0.0329,0.0106,0.0049,-0.0011
5,0.0052,0.0194,0.0389,0.0860,0.1065,0.0913,0.0541,0.0056,0.0136,0.0114,0.0038,-0.0010
6,-0.0011,0.0020,0.0016,0.0247,0.0913,0.1986,0.2193,0.0579,-0.0522,0.0092,0.0021,-0.0009
7,-0.0034,-0.0037,-0.0229,-0.0277,0.0541,0.2193,0.3002,0.1147,-0.0537,0.0126,0.0008,-0.0009
8,-0.0006,0.0048,-0.0096,-0.0279,0.0056,0.0579,0.1147,0.1296,0.0645,0.0100,-0.0004,-0.0009
9,0.0079,0.0269,0.0288,0.0329,0.0136,-0.0522,-0.0537,0.0645,0.1420,0.0151,0.0031,-0.0004
10,0.0020,0.0068,0.0062,0.0106,0.0114,0.0092,0.0126,0.0100,0.0151,0.0085,0.0011,-0.0002
11,0.0008,0.0022,0.0031,0.0049,0.0038,0.0021,0.0008,-0.0004,0.0031,0.0011,0.0014,0.0000
12,-0.0001,-0.0004,-0.0008,-0.0011,-0.0010,-0.0009,-0.0009,-0.0009,-0.0004,-0.0002,0.0000,0.0003
\end{filecontents*}
\csvautotabular{crossSectionCovariance.csv}
 \caption{Covariance matrix for the data cross section shown in Fig. \ref{fig:crossSectionMnvTune}.  All values in this table are to be scaled by $10^{-78}$~$cm^{4} \times c^{2} / GeV^{2} / nucleon^{2}$.  The error bars on Fig. \ref{fig:crossSectionMnvTune} are the square root of the diagonal terms of this matrix.  Underflow and overflow bins are not included.}
\end{table*}

\section{Selection Efficiency}
\begin{figure}[bhtp]
  \centering
  \includegraphics[scale=0.4]{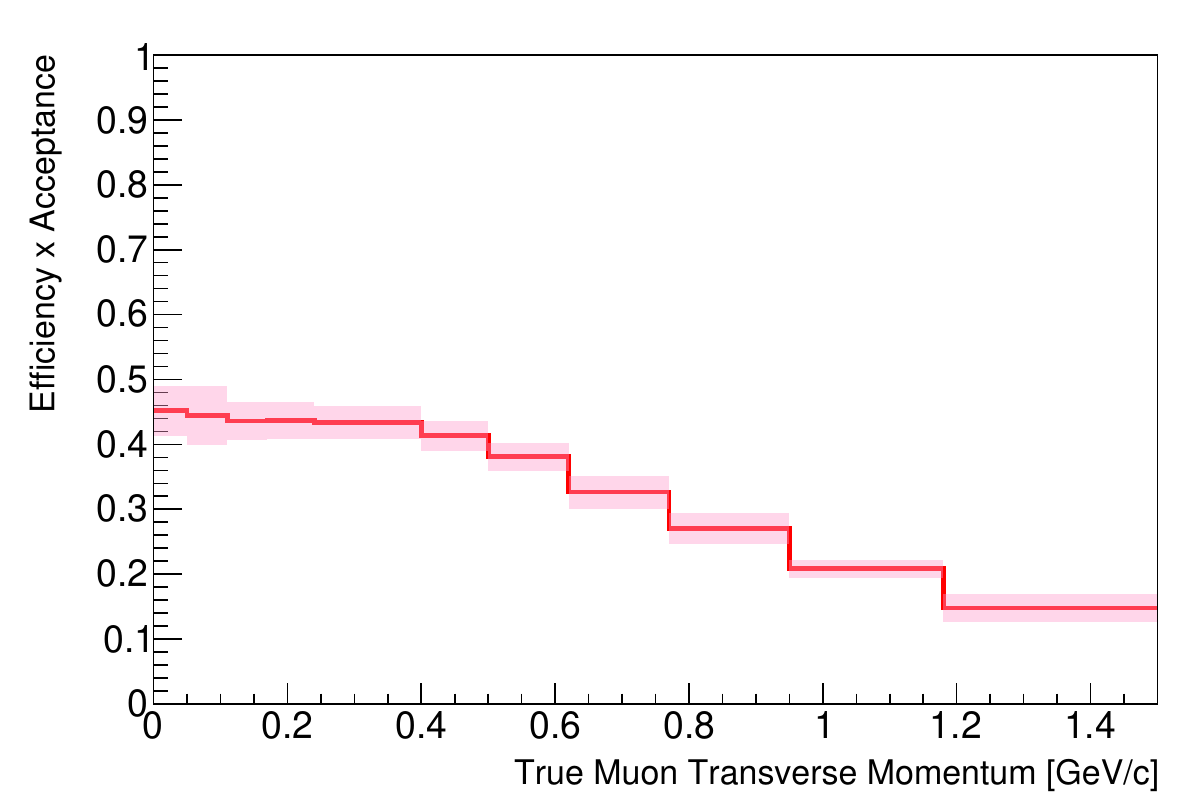}
  \caption{Efficiency to select a low $E_{available}$ multi-neutron CC antineutrino interaction as a function of muon transverse momentum.  This includes effects from muon selection, the $E_{available}$ cut, and the neutron reconstruction efficiency shown in Figure \ref{fig:neutronDetectionEfficiency}.  The pink band shows the combined statistical and systematic uncertainties on efficiency.  Efficiency is expected to gradually drop off a high $p_{T \mu}$ because this region has many muons that miss MINOS when they leave the MINERvA tracker.}
  \label{fig:efficiency}
\end{figure}

\end{document}